\documentclass{PoS}
\usepackage{amsmath,amsfonts,amssymb}
\usepackage{color}

\def\LP{\left(}         
\def\RP{\right)}        

\def\PARTWO#1#2{ {{\partial^2 #1}\over{\partial #2}^2} }

\newcommand{\BE}{\begin{displaymath}}
\newcommand{\EE}{\end{displaymath}}
\def\BNE{\begin{equation}}
\def\ENE{\end{equation}}
\def\BEA{\begin{eqnarray}}
\def\EEA{\nonumber\end{eqnarray}}
\def\EEAN{\end{eqnarray}}
\def\EL{\nonumber\\}
\def\chpt{$\chi$PT}
\def\rcite#1{Ref.~\cite{#1}}
\def\Rcite#1{Reference~\cite{#1}}

\def\Rcites#1{References~\cite{#1}}
\def\eqn#1{\label{eq:#1}}

\def\eq#1{Eq.~(\ref{eq:#1})}

\title{Non-equilibration of topological charge and its effects}

\ShortTitle{Non-equilibration of topology}

\author{Claude Bernard\\
        Physics Department, Washington University, Saint Louis, MO 63130, USA\\
        E-mail: \email{cb@lump.wustl.edu}}

\author{\speaker{Doug Toussaint}\\
        Physics Department, University of Arizona, Tucson, AZ 85721, USA\\
        E-mail: \email{doug@physics.arizona.edu}}

\abstract{
In QCD simulations at small lattice spacings, the topological charge $Q$
evolves very slowly and, if this quantity is not properly equilibrated,
we could get incorrect results for physical quantities, or incorrect
estimates of their errors.  We use the known relation between the dependence
of masses and decay constants on the QCD vacuum angle $\theta$ and the
squared topological charge $Q^2$  together with chiral perturbation theory
results for the dependence of masses and decay constants on $\theta$ to
estimate the size of these effects and suggest strategies for dealing
with them. For the partially quenched case, we sketch an alternative
derivation of the known $\chi$PT results of Aoki and Fukaya, using the
nonperturbatively correct chiral theory worked out by Golterman, Sharpe and
Singleton, and by Sharpe and Shoresh. With the MILC collaboration's ensembles of lattices with four
flavors of HISQ dynamical quarks, we measure the $Q^2$ dependence of masses
and decay constants and compare to the $\chi$PT forms.  The observed
agreement gives us confidence that we can reliably estimate the
errors from slow topology change, and even correct for its
leading effects.
          }

\FullConference{34th annual International Symposium on Lattice Field Theory\\
		24-30 July 2016\\
		University of Southampton, UK}

\begin{document}

\section{Evolution of $Q$}


QCD simulations using (approximately) continuous evolution algorithms show very
slow evolution of the topological charge $Q$ when the lattice spacing is small.
This is expected,
since in  the continuum theory $Q$ cannot change in a
continuous evolution of the fields.  This is a concern for QCD
simulations since the distribution of $Q$ may not be properly
sampled in a simulation of practical length.  Here we study the evolution of
$Q$ in the MILC collaboration's ensembles of lattices
with a one-loop Symanzik and tadpole improved gauge action and four flavors
of highly improved staggered quarks (HISQ).  We use chiral perturbation
theory to compute the effects of poor sampling of $Q$ on pseudoscalar masses
and decay constants, and compare these results to our simulations.

\begin{figure}
\vspace{-1.30in} \hspace{-0.25in}
\includegraphics[width=1.05\textwidth]{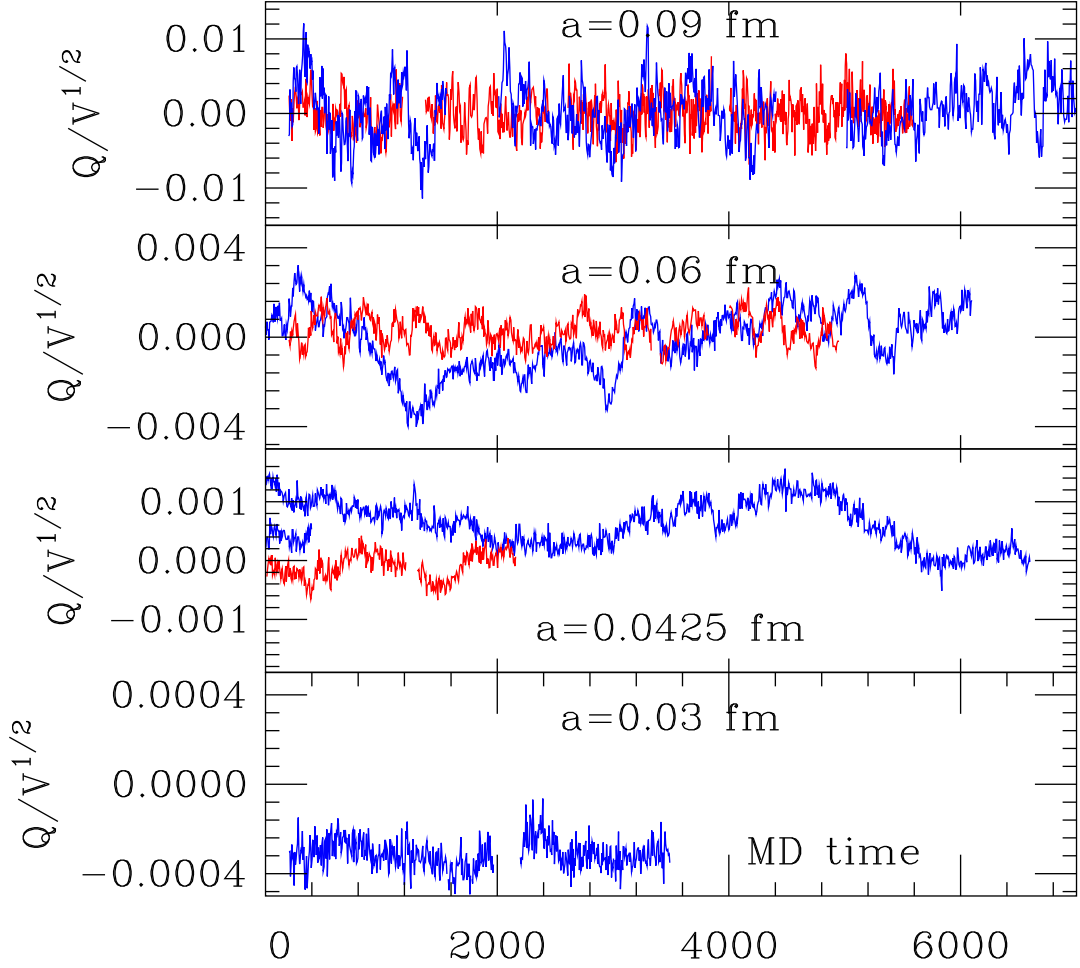}
\vspace{-2.20in}
\caption{
\label{topo_history_fig}
Topological charge time histories for various lattice spacings.  Blue traces are for ensembles
with light sea quark mass one fifth of the strange quark mass and red traces for ensembles with
light sea quark mass at its physical value.
Notice the narrower distributions and shorter autocorrelation times for physical quark mass
ensembles. Multiple traces in some graphs correspond to multiple runs, sometimes with differing trajectory lengths.
}
\end{figure}

The ensembles we study have lattice spacings ranging from $0.09$ fm to $0.03$ fm,
and light sea quark masses either at one fifth of the strange quark mass or
approximately tuned to the physical light quark mass.
Figure~\ref{topo_history_fig} shows the time histories of $Q/V^{1/2}$, where
$V$ is the four dimensional lattice volume in fm$\null^4$.  In this plot the blue
traces are for ensembles
with light sea quark mass one fifth of the strange quark mass and red traces for ensembles with
physical light quark mass.  The increasing autocorrelation time of $Q$ as $a$
gets small is clearly visible, and we see that at $a=0.03$ fm the simulation has not yet
explored most of the desired values of $Q^2$.
We also see that for each lattice spacing the local structure of the time histories is very similar
for the $m_l=m_s/5$ ensemble and the physical $m_l$ ensemble.  However, in the
$m_l=m_s/5$ ensemble $Q$ ranges over larger values, therefore taking longer to random walk
through this range, leading to a longer autocorrelation time.  This is as
expected, since the gauge action controls the tunneling rate for $Q$, so
the average squared change in $Q$ per unit volume per unit simulation time is approximately
independent of the light quark mass.   However, the fermion determinant does suppress the
average $Q^2$, and we expect  the topological susceptibility, $\langle Q^2/V \rangle$,
to be approximately proportional to $m_l$.

\begin{figure}
\vspace{-1.35in} 
\begin{center}\includegraphics[width=.70\textwidth]{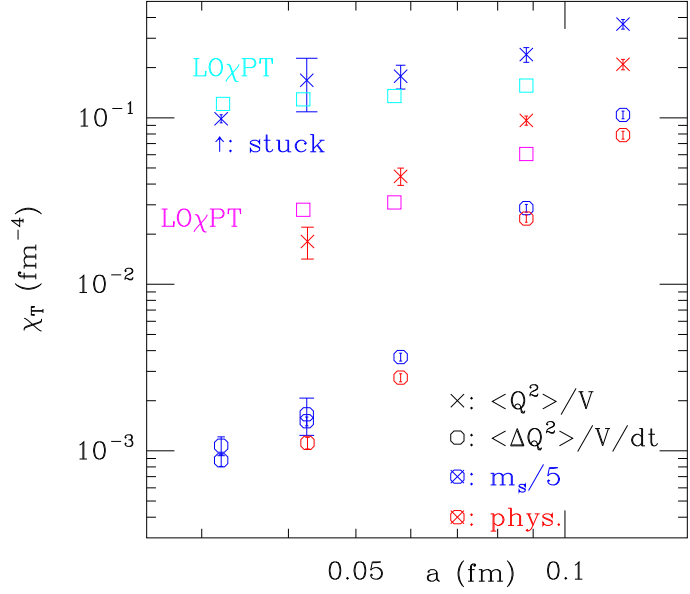}\end{center}
\vspace{-1.35in}
\caption{
\label{qsq_vs_a_fig}
Average topological susceptibility $<Q^2/V>$ (crosses) and tunneling rate 
$\langle (\Delta Q)^2 \rangle$ (octagons) versus lattice spacing.
The cyan and magenta squares are the lowest order chiral perturbation theory
predictions for the susceptibility.
}
\end{figure}

Figure~\ref{qsq_vs_a_fig} shows the tunneling rate, $\langle \LP \Delta Q \RP^2/\LP V dt \RP \rangle$
with octagons, where the blue symbols are for the $m_s/5$ ensembles and the red for the physical
$m_l$ ensembles.  We see that the tunneling rate doesn't depend much on the quark
mass, but is decreasing as expected as $a$ gets small. (In the cases where there are two blue
octagons, there were two sub-ensembles with a different molecular dynamics trajectory length in
each sub-ensemble.)
The crosses in Fig.~\ref{qsq_vs_a_fig} show the topological susceptibility, $\langle Q^2/V \rangle$.
Here we see the expected strong dependence on light quark mass.  The small error bar
on the $0.03$ fm point is unrealistic --- it simply reflects the fact the $Q$ is basically stuck
near this value in this simulation.

\section{Theoretical treatment of the dependence on topological charge}

The topological susceptibility is defined by \cite{leutwylersmilga1992}
\BEA Z(\theta) &=& \int {\cal D}A {\cal D}\bar\Psi {\cal D}\Psi \
\exp(- S[A,\bar\Psi,\Psi]) \exp(- i \theta Q[A])\eqn{Z}\\
\chi_t &\equiv& -\frac{1}{V} \; \left(\frac{1}{Z}\frac{\partial^2 Z}{\partial\theta^2}\right)\Bigg\vert_{\theta=0}  = \ \frac{1}{V} \langle Q^2\rangle .\eqn{chi}\EEAN
Quantities evaluated at fixed $Q$ are found by Fourier transforming
\BEA Z_Q &=&\frac{1}{2 \pi} \int_{-\pi}^\pi d\theta \ \exp(i \theta Q) Z(\theta) \EL
G_Q &=& \langle {\cal O}_1 {\cal O}_2 ... {\cal O}_n \rangle_Q =
\frac{1}{Z_Q} \frac{1}{2 \pi} \int_{- \pi}^\pi d\theta \
  \exp(i \theta Q) Z(\theta) G(\theta)  \EEA
with $G(\theta)= \langle {\cal O}_1 {\cal O}_2 ... {\cal O}_n \rangle_\theta$.
For large 4-dim volumes $V$, we can do the $\theta$ integrals by the saddle point method
to find  (for $B$ the mass $M$ or the decay constant $f$) \cite{brower2003,aoki2007}
\BNE
\label{topo_dep_eq}
B \big|_{Q,V} = B + \frac{1}{2\chi_T V} B''
   \LP 1-\frac{Q^2}{\chi_T V} \RP + {\cal O}\LP \frac{1}{\LP \chi_T V\RP^2} \RP \ENE
where  $B'' \equiv \PARTWO{B}{\theta}\big|_{\theta=0}$.  By \eq{chi},
the correction vanishes when averaged over $Q$.

Since the quantities $M''$ and $f''$ are physical, we can get a theoretical handle on topological effects 
by calculating
them in continuum, infinite volume, \chpt. A first calculation of $M''$ in \chpt\ for full (unitary) QCD appears in \rcite{brower2003}.  Since most
of our lattice data is partially quenched, we need to extend the calculation to $M''$ and $f''$ to 
partially quenched \chpt\ (PQ\chpt).   \Rcite{aokifukaya2009}  worked this out, using 
the replica method to remove the determinant of the valence quarks.
However, the required calculation is non-perturbative, at least on its face, since  the vacuum state changes 
in the presence of $\theta$.  The replica method is only justified perturbatively, so a non-perturbatively safe
method is preferable. The
Lagrangian approach of \rcite{bernardgolterman1993}, which introduces ghost (bosonic) quarks to 
cancel the valence quark determinant, is also only valid perturbatively, since it ignores the 
requirement that  bosonic path integral be convergent.  
\Rcites{gss,sharpeshoresh2001} fixed the non-perturbative problems of the Lagrangian approach by taking into account the convergence requirement.

The PQ\chpt\ Lagrangian in the presence of $\theta$  for $n_F$ sea quarks, $n_V$ valence quarks, and $n_V$ ghost quarks 
is \cite{gss,sharpeshoresh2001}  
\vspace{-4mm}
\begin{equation}
{\cal L}= \frac{f^2}{8}{\rm str}(\partial_\mu\Sigma\partial_\mu\Sigma^{-1})
-\frac{Bf^2}{4}{\rm str}(e^{-i\theta/n_F}{\cal M}\Sigma + e^{i\theta/n_F}{\cal M}\Sigma^{-1}),
\vspace{-1mm}
\end{equation}
where str is the supertrace, the factors $e^{\pm i\theta/n_F}$ arise from an anomalous rotation to remove 
the $i\theta Q$ term in \eq{Z}, and 
$\Sigma$ is an  $(n_F+2n_V)\!\times\!(n_F+2n_V)$ matrix constructed from
the meson field $\Phi$:
\vspace{-1.5mm}
\begin{equation}\eqn{Phi}
 \Sigma=e^{2i\Phi/f},\qquad \Phi=
\begin{pmatrix} \phi & \bar\chi \\ 
                         \chi & -i\hat \phi 
\end{pmatrix}.
\vspace{-1.5mm}
\end{equation}
Here $\phi$ and $\hat\phi$ are hermitian bosonic fields%
\footnote{Technically, this applies to the ``body'' of $\hat\phi$.},
representing quark-quark and ghost-ghost mesons, 
respectively.  The quark-ghost fields $\chi$ and $\bar\chi$ are fermionic.
The field $\hat\phi$ is integrated from $-\infty$ to $+\infty$; the factor $i$ in \eq{Phi} ensures a convergent 
path integral, making the  $\hat\phi$ action positive definite.  Convergence for
the $\phi$ integral is not a problem because the domain of $\phi$ is a compact space, as usual. 
Subtleties for fields along the diagonal have been ignored in \eq{Phi} for simplicity.

In the full theory, we would now minimize the potential energy term to find  the vacuum state
$\langle \Sigma \rangle$.  Here, the potential energy is complex. \Rcite{gss} argues that  we should 
therefore find a saddle point  (deforming the $\hat\phi$ contour as needed), not a minimum.
We note that, unlike what happens in  the quenched case 
\cite{gss},  the symmetry between valence and ghost quarks is automatically preserved by the
saddle point, and does not need to be imposed by hand.
To find the $\theta$-dependence of the mass (at tree level), we can then expand 
the Lagrangian to quadratic order 
in $\Phi$ around the vacuum state by writing
\vspace{-1.5mm}
\begin{equation}
\Sigma = \sqrt{ \langle \Sigma \rangle}\; e^{2i\Phi/f}  \sqrt{\langle \Sigma \rangle}.
\end{equation}
 This way of expanding keeps ``extended parity'' symmetry (parity + $\theta\!\to\!-\theta$) simple:  $\Phi\to
-\Phi$, $\Sigma \to \Sigma^{-1}$.   
For the decay constant, we similarly expand the axial current to linear order in $\Phi$.

For $n_F=3$ with masses $m_u=m_d=m$ and $m_s$, and $n_V=2$ with masses $m_x$, $m_y$, we find
\vspace{-1.5mm}
\begin{eqnarray}
\label{chiptresults_eq}
M_{xy}'' &=& -M_{xy}\;\frac{m^2 m^2_s}{2(m+2m_s)^2}\;\frac{1}{m_x m_y}, \EL
f_{xy}'' &=&  -f_{xy}\; \frac{m^2m_s^2}{4(m+2m_s)^2}\;\frac{(m_x-m_y)^2}{m^2_x m^2_y} .
\vspace{-3mm}
\end{eqnarray}
These results agree with those in \rcite{aokifukaya2009}.  This indicates that 
it is not necessary to use a nonperturbatively
correct approach for this problem; the reason seems to be
that in the end we only need the new vacuum state in the infinitesimal neighborhood of $\theta=0$.
The singular limit as $m_x$ or $m_y\to0$ presumably comes 
from topological zero modes, which are not suppressed by low valence-quark mass since the 
valence determinant is absent.   
For $n_F=4$, decoupling works (if $m_c$ is sufficiently heavy), so we can still use the above results.

\begin{figure}
\vspace{-0.81in}
\begin{center}\includegraphics[width=.6\textwidth]{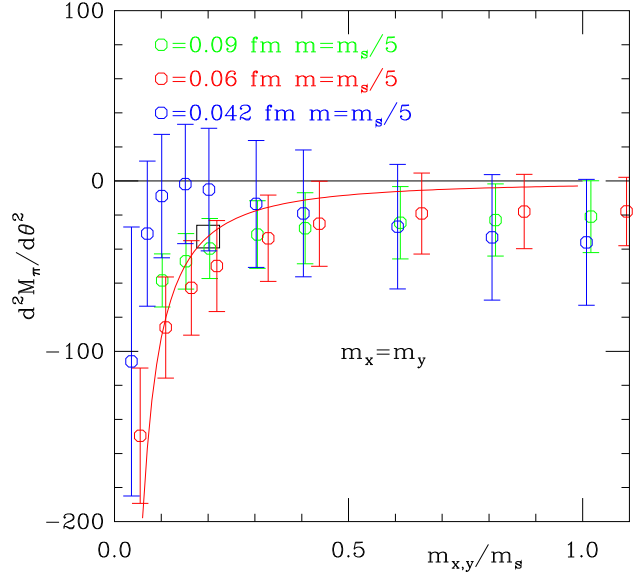}\end{center}
\vspace{-1.15in}
\caption{
\label{d2m_fig}
$\PARTWO{M}{\theta}$ on ensembles with $m_l=m_s/5$ along the line $m_x=m_y$.
The red line is the PQ$\chi$PT prediction (no free parameters), and the square
is the unitary point.
}
\end{figure}

\section{Comparison to simulation results}

We have calculated pseudoscalar masses and decay constants on the HISQ ensembles described above, using
methods described in Ref.~\cite{fermimilcdecay2014}.  To find the dependence on the
topological charge, we use the results of a single-elimination jackknife analysis of these
quantities together with the time histories of topological charge shown above.  We first construct
effective masses and decay constants for each lattice by taking the ensemble average
values minus $N$ times the deviation of the corresponding jackknife sample value from the
ensemble average.  Then we fit to a linear function of the topological charge, $M = M_0 + \frac{C}{2}Q^2$,
 assigning each
data point an error equal to the standard deviation of the distribution.
(Strictly speaking, it should be the standard deviation reduced by the contribution of
the dependence on topological charge to the variance, but this turns out to make little
difference.)  
Then we use Eq.~\ref{topo_dep_eq} to convert $C$ into a derivative with respect to $\theta$.

The results are noisy, but consistent with the $\chi PT$ predictions.  Statistically
significant signals are found in the $m_l=m_s/5$ ensembles, since these have much smaller
physical volumes than the physical light quark mass ensembles.  For example, in the $a\approx 0.06$ fm
ensembles the $m_l=m_s/5$ lattices have a volume of $180$ fm$\null^4$, while the physical
$m_l$ lattices have a volume of $1920$ fm$\null^4$.  Also, Eq.~\ref{chiptresults_eq} shows
that the derivatives of the masses and decay constants have a partially quenched divergence
when $m_x$ or $m_y$ goes to zero with $m_l$ fixed, and for the $m_l=m_s/5$ ensembles we
have used valence quark masses smaller than $m_l$, in some cases as small as the
physical $m_l$.

\begin{figure}
\vspace{-1.0in}
\begin{center}\begin{tabular}{ll}
\hspace{-0.4in} \includegraphics[width=.57\textwidth]{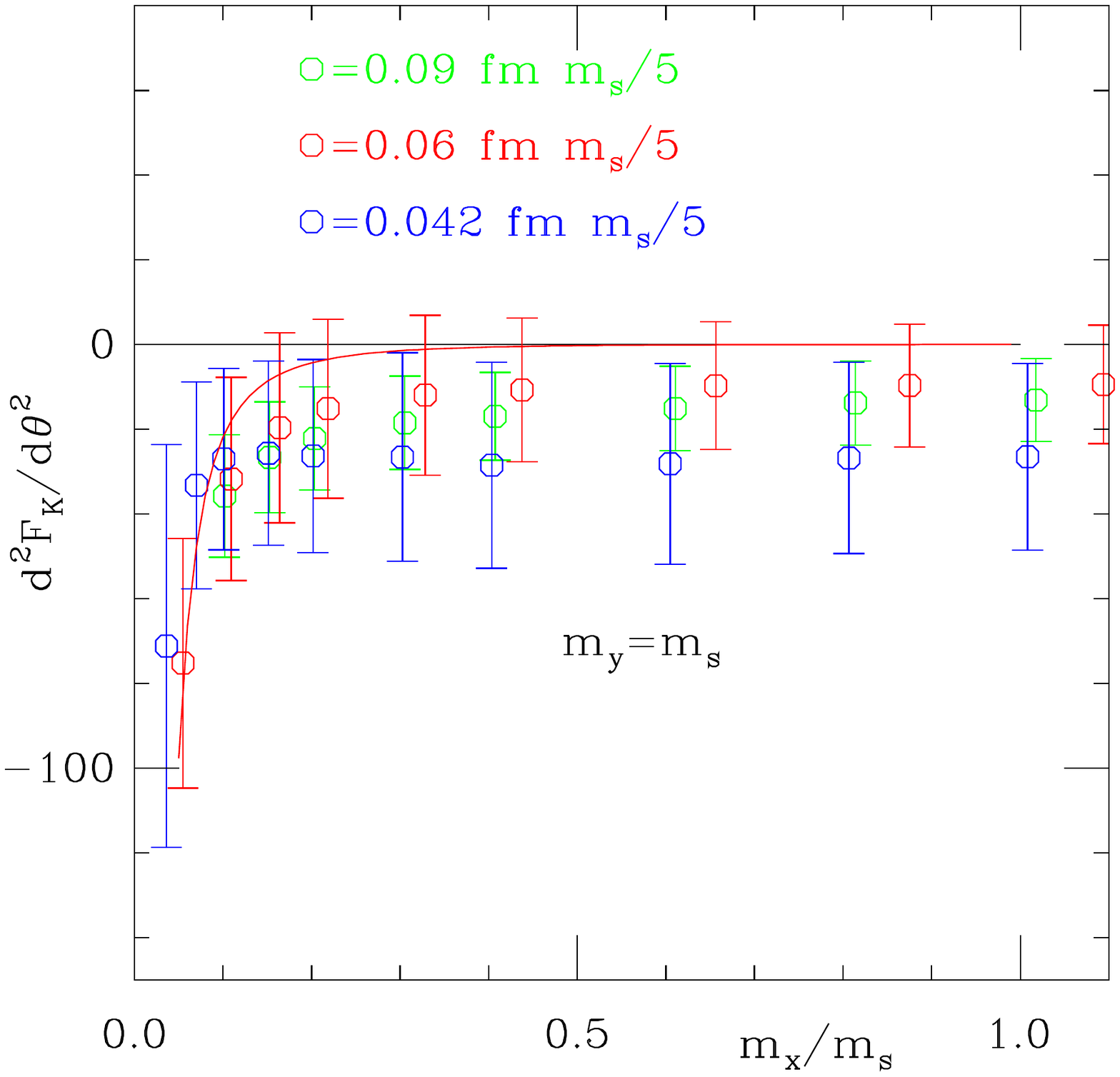} &
\hspace{-0.5in} \includegraphics[width=.57\textwidth]{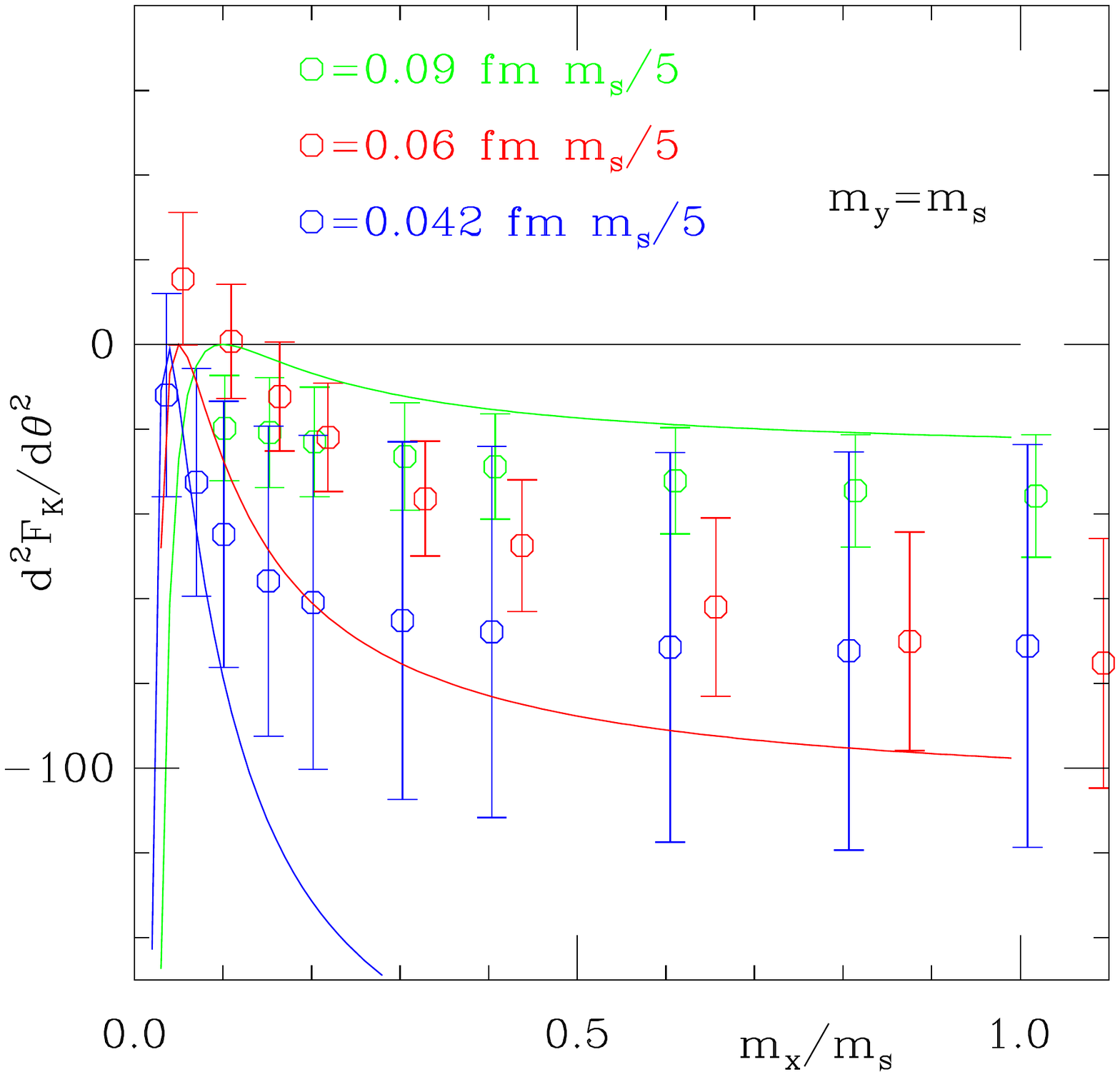} \\
\end{tabular}\end{center}
\vspace{-1.05in}
\caption{
\label{d2f_fig}
$\PARTWO{F}{\theta}$ on ensembles with $m_l=m_s/5$.
The left panel shows $\PARTWO{F}{\theta}$ as a function of one valence
quark mass, $m_x$ along the line $m_y=m_s$.
The red line is the PQ$\chi$PT prediction (no free parameters), which vanishes for degenerate quarks.
The right panel shows the quantity  with $m_y$
fixed to the smallest available value.
The lines are the PQ$\chi$PT predictions.  There are three separate lines
in the right panel because the smallest valence quark mass was different in each ensemble,
$0.1\, m_s$, $0.05\,m_s$ and $0.037\,m_s$ for the $0.09$, $0.06$ and $0.042$ fm ensembles
respectively.
}
\end{figure}

Figure~\ref{d2m_fig} shows $\PARTWO{M}{\theta}$ for the $m_s/5$ ensembles for degenerate valence quark
masses, $m_x=m_y$.  The red line in the figure is the $PQ\chi PT$ prediction in Eq.~\ref{chiptresults_eq},
which we emphasize is a prediction with no free parameters.
Obviously the statistical errors are large, but they are consistent with the prediction, and the
divergence at small valence quark mass is clearly seen.
Since $\PARTWO{F}{\theta}$ vanishes for degenerate valence quarks, we plot this quantity
along different lines in Fig.~\ref{d2f_fig}.  The left panel shows $\PARTWO{F}{\theta}$
as a function of one valence quark mass, $m_x$, with the other fixed at the strange quark
mass, together with the $\chi PT$ prediction.  The right panel shows $\PARTWO{F}{\theta}$
along lines where $m_y$ is held fixed at the lightest valence quark mass available in
each ensemble.  The vanishing of $\PARTWO{F}{\theta}$ when the valence quarks are
degenerate is particularly striking in this plot.

Knowing the dependence of masses and decay constants on the average $Q^2$, we can correct
our simulation results to account for the difference of the average in our simulation, $<Q^2>_{sample}$
and the correct $<Q^2>$.  To estimate this correct $<Q^2>$ we use the lowest order $\chi PT$ result,
$\chi_T = \frac{f_\pi^2}{4} \overline{M_I^2}$ where $1/\overline{M_I^2} = 2/M_{\pi,I}^2 + 1/M_{ss,I}^2$ 
\cite{leutwylersmilga1992}.
Here the ``$I$'' indicates the taste singlet masses \cite{tastesinglettopo}.
The $\chi PT$ results are shown in Fig.~\ref{qsq_vs_a_fig}.  For large $a$ the deviation from
the lowest order $\chi PT$ results is due to lattice artifacts, probably mostly higher order
taste breaking effects, but for $a=0.042$ and $0.03$ fm we expect the $\chi PT$ results to be
pretty good.  For an example of the size of these effects in our simulation, we look at $f_K/f_\pi$
in our two ensembles with $a \approx 0.042$ fm.  This ratio has very small statistical errors, so this
is a good test.
To make this correction, rearrange Eq.~\ref{topo_dep_eq} as
\BNE f_{corrected} = f_{sample} - \frac{1}{2\chi_T V} F^{\prime\prime} \LP 1 - \frac{<Q^2>_{sample}}{\chi_T V} \RP \ENE
For the $0.042$ fm physical $m_l$ ensemble, with $L=6.05$ fm and estimating $\frac{<Q^2>_{sample}}{\chi_T V} \approx 0.7$
we find a fractional shift $\frac{\Delta f}{f} = 0.0002$.  This can be compared to our statistical error
on this ratio, $0.0010$ and to the ``conventional'' finite size effects from pions propagating around the
periodic lattice, estimated in NLO 
staggered $\chi PT$, of 0.0009. The effects are 
larger in the ensemble with $m_l/m_s=0.2$, since these lattices have
much smaller volume and a partial quenching divergence.  In this case a similar estimate gives $\frac{\Delta f}{f} \approx -0.002$ to be
compared with a statistical error of $0.003$.

We close by noting that this strategy is in the same spirit as our treatment of ``conventional''
finite size effects.  We use $\chi PT$ to estimate the effects and correct our results, and estimates
of the effects of higher order $\chi PT$ and/or uncertainties in the $\chi PT$ parameters should be
included in the systematic error budget.

\vspace{-0.2in}
\section*{Acknowledgements}
\vspace{-0.1in}

This work was supported by US DOE contracts DE-FG02-91ER40628 and DE-FG02-13ER-41976.
Computations were done at centers supported by the US DOE and NSF, including ALCF, NCSA Bluewaters,
NERSC, TACC, NICS, NCAR and USQCD facilities.

\end{document}